\documentclass[
sigconf
]{acmart}

\AtBeginDocument{
  \providecommand\BibTeX{{%
    \normalfont B\kern-0.5em{\scshape i\kern-0.25em b}\kern-0.8em\TeX}}}

\copyrightyear{2023}  
\acmYear{2023}  
\setcopyright{rightsretained}  
\acmConference[EAAMO '23]{Equity and Access in Algorithms, Mechanisms, and Optimization}{October 30-November 1, 2023}{Boston, MA, USA} 
\acmBooktitle{Equity and Access in Algorithms, Mechanisms, and Optimization (EAAMO '23), October 30-November 1, 2023, Boston, MA, USA}
\acmDOI{10.1145/3617694.3623237} 
\acmISBN{979-8-4007-0381-2/23/11}

\usepackage{makecell}
\usepackage{colortbl}

\usepackage[suppress]{color-edits}
\addauthor [Ben] {bl} {blue}

\begin{document}

\title[Strategic Evaluation]{Strategic Evaluation: Subjects, Evaluators, and Society}

\author{Benjamin Laufer}
\email{bdl56@cornell.edu}
\affiliation{%
  \institution{Cornell Tech}
  \city{New York}
  \state{New York}
  \country{USA}
}

\author{Jon Kleinberg}
\email{kleinberg@cornell.edu}
\affiliation{%
  \institution{Cornell University}
  \city{Ithaca}
  \state{New York}
  \country{USA}
}

\author{Karen Levy}
\email{karen.levy@cornell.edu}
\affiliation{%
  \institution{Cornell University}
  \city{Ithaca}
  \state{New York}
  \country{USA}
}

\author{Helen Nissenbaum}
\email{hn288@cornell.edu}
\affiliation{%
  \institution{Cornell Tech}
  \city{New York}
  \state{New York}
  \country{USA}
}

\begin{abstract}

A broad current application of algorithms is in formal and quantitative measures of murky concepts -- like merit -- to make decisions. When people strategically respond to these sorts of evaluations in order to gain favorable decision outcomes, their behavior can be subjected to moral judgments. They may be described as `gaming the system' or `cheating,' or (in other cases) investing `honest effort' or `improving.' Machine learning literature on strategic behavior has tried to describe these dynamics by emphasizing the efforts expended by decision subjects hoping to obtain a more favorable assessment --- some works offer ways to preempt or prevent such manipulations, some differentiate `gaming’ from `improvement’ behavior, while others aim to measure the effort burden or disparate effects of classification systems. 

We begin from a different starting point: that the design of an evaluation 
\textit{itself} can be understood as furthering goals held by the evaluator which may be misaligned with broader societal goals.
To develop the idea that evaluation represents a strategic interaction in which both the evaluator and the subject of their evaluation are operating out of self-interest,
we put forward a model that represents the process of evaluation using three interacting agents: a decision subject, an evaluator, and \textit{society}, representing a bundle of values and oversight mechanisms. We highlight our model's applicability to a number of social systems where one or two players strategically undermine the others' interests to advance their own. Treating evaluators as themselves strategic allows us to re-cast the scrutiny directed at decision subjects, towards the incentives that underpin institutional designs of evaluations.
In practice, the moral standing of strategic behaviors often depend on the moral standing of the evaluations and incentives that provoke such behaviors. We apply our framework to a variety of extended examples and discuss ethical implications. 
  
\end{abstract}

\keywords{Strategic behavior, Evaluation, Measurement}

\maketitle

\section{Introduction}

An important recent theme in machine learning and mechanism design research has been its expansion into domains such as hiring, lending, educational admissions, and other settings where people apply for opportunities (a job, a loan, a place at a selective college) and an institution draws on algorithmic assistance for evaluating them. Because of the increasing interest in such applications, a growing body of work focuses on the strategic behaviors of people subjected to these types of algorithmic decisions---that is, attempts to modify one’s attributes to obtain a more favorable classification in an algorithmic evaluation \cite{bambauer2018algorithm}. Recent work on this issue has explored a range of different bases for such strategic behavior, ranging from strategies that improve the attributes of interest in the evaluation (such as when a student studies the material on a test so as to score higher) to strategies that ``game'' the process by improving measurable features without necessarily affecting the attributes the evaluation is designed to measure (such as when a student devotes time to perfecting test-taking strategies, rather than mastering the underlying material).

The work on these issues in the computer science and mechanism design communities has thus conceived of the process as a two-player game between an evaluator and a decision subject, with the institution constructing an evaluation designed to robustly measure certain attributes of the decision subject, and the decision subject investing effort to score well on the evaluation. This two-player structure, in which the first party engages in self-presentation while the second party focuses on eliciting underlying information about them, is indeed so central to the work in this area that it is often not explicitly called out as an assumption of the models being used. But if we think about the issues that arise in the process of evaluation, it is clear that some of them are not well-explained by this type of two-player interaction. For example, sociological and organizational research has demonstrated that evaluating candidates according to their ``fit'' with company culture can encode underlying cultural biases and cement inequities in an organization. If a job applicant spends energy presenting themselves in a way that conveys fit in this sense, should we think of them as ``gaming'' the hiring process? Or, alternatively, should we think of the hiring process itself as deficient in some way? A model that treats the evaluation merely as an elicitation device for an applicant’s attributes will struggle to identify the deeper normative concerns at play in such an example.

Consider a second example: when a university assigns grades to its students, we typically describe this grading process as a way of measuring student performance. But in this narrow view, we may miss some of the other interests at play in a grading scenario, together with their normative implications.  A university that is engaging in grade inflation, for instance, might find that its instructors receive higher teaching evaluations and its development program receives larger alumni donations in a regime with higher grades. Students may also view themselves as benefitting from such a regime. If both parties appear to be advantaged by the decision to assign uniformly high grades, how do we pinpoint what is intuitively undesirable about grade inflation? 

These and many other examples suggest that a fuller understanding of strategic evaluation requires that we include an additional player in the model. We draw on sociological theory and empirical evidence about how evaluating institutions function in society: while evaluating institutions are frequently tasked, explicitly or implicitly, with implementing broad societal goals and values, they also operate out of self-interest, which may be more or less aligned with these societal aims. From this starting point, we observe that in a richer model of strategic evaluation, it is not only the decision subjects who operate out of self-interest: the design of an evaluation should \textit{itself} be understood as a self-interested behavior by the institution, which aims to achieve its own goals under various social, legal, and organizational constraints. The institution is in turn held to account by a third player, which we can think of as playing the role of society---in the form of laws, regulations, norms, or individual authorities tasked with oversight.  

Our introductory examples suggest that many of the central considerations in the design of evaluations are better understood as clashes within this three-player structure, between the strategic actions of the individuals being evaluated, the institution performing the evaluation, and society’s expectations for what the evaluation should be achieving. 

The paper is organized as follows. Section \ref{overview} provides our three-player model, aiming to capture the various ways that an evaluation outcome can diverge from societal goals. Section \ref{sec:examples} discusses three extended examples: hiring practices, grade inflation, and sports. In Section \ref{sec:mechanical}, we use the model to enumerate the set of possible scenarios where the interests of the three players either align or diverge. Section \ref{sec:ethical} discusses the ethical implications of our model, aiming to recast ethical scrutiny in light of evaluators' strategic aims. We discuss further related work in Section \ref{sec:related}.

\section{Overview}\label{overview}

We suppose that society has determined some desirable property of interest, and it would like to find the people who exhibit this property.
But since society lacks the ability to perform this assessment itself,
it delegates the task to an {\em evaluator}.
The evaluator constructs a test that it gives to {\em subjects},
and those who pass the evaluation are deemed to satisfy the property. (``Society,'' of course, is not a monolithic actor with a single set of goals. As our model will illustrate, we intentionally conceive of society capaciously---encompassing both situations in which a governing body is imbued with some regulatory authority to oversee the activities of evaluators, as well as situations in which societal interests are manifested as the expression of public values, but without explicit organizational oversight.)

There are several sources of `slippage' that are possible in this setting---that is, cases in which an assessment fails to meet its mandate or achieve broader societal goals. We would like a model that is capable of considering these discrepancies in a unified manner.
The first source of slippage is the gap between someone's performance on a test and their underlying ability. A subject might have slept badly the night before a math test, leading to a score that does not reflect their skills. Or, a strategic test-taker might have chosen to invest in skills that boost their score on the assessment without improving underlying properties (i.e., some form of `gaming').
A second source of slippage is the gap between the aim of the evaluator and the design of the evaluation. No evaluation perfectly measures its intended quantity. Myriad factors constrain the evaluator and limit the signal that an evaluation can capture, either by introducing noise or bias to the measurement. This type of misalignment has received significant attention in 
work on the role of {\em proxies} in classification: practical
evaluations generally need to rely on measurable properties that
stand in for the property of interest, but which do not precisely
coincide with it.
The third source of slippage is the gap between the evaluator's aim and the true property of interest to society. Once we appreciate that the evaluator, like the subject, is also
a strategic actor with their own self-interest, then we can see that the misalignment 
can also arise from forms of gaming or cheating by the evaluator:
the evaluator might care about a property other than the one
society is interested in assessing, 
and they might have correspondingly
created a test that is better at measuring
their property than it is at measuring the one of
interest to society.

\bldelete{
There are several sources of slippage that are possible in this setting,
and we would like a model that is capable of considering all of them
in a unified setting.}
\bldelete{
The first source of slippage is the gap between someone's performance on
a test and what the test is intended to measure.
This type of misalignment can arise from forms of {\em strategic behavior}
by the subject, such as gaming or cheating; it can also arise
from more straightforward implementation failures in the
design of the test.

The second source of slippage is the gap between what the test
is intended to measure and the true property of interest.
This type of misalignment has received significant attention in 
work on the role of {\em proxies} in classification: practical
evaluations generally need to rely on measurable properties that
stand in for the property of interest, but which do not precisely
coincide with it.
Once we appreciate that the evaluator, like the subject, is also
a strategic actor with their own goals, then we can see that the misalignment 
can also arise from forms of gaming or cheating by the evaluator:
the evaluator might care about a property other than the one
society was initially interested in assessing, 
and they might have correspondingly
created a test that is better at measuring
their property than it is at measuring the one of
interest to society.
}

There are thus multiple issues that we need to keep in focus
for the purposes of our analysis: a given circumstance inhabited by
the subject might or might not be sufficient to pass a test;
a given way of passing the test might or might not correspond
to what the evaluator is trying to measure;
and what the evaluator is trying to measure might or might not
correspond to the underlying societal values that motivated the
test in the first place.
To keep track of these issues we therefore introduce a formal model
that includes all of these facets.
The model cannot by itself resolve the underlying ethical questions
about the behavior of the subject and the evaluator in any
given situation, but it can provide precision about the nature
of these situations as a starting point for ethical analysis.

\subsection*{The Model}
\label{subsec:model}

In order to formalize the notion that the test is imperfectly
measuring an underlying property of interest, we need to represent
the idea that the true state of the subject is hidden and only
partially observable.
Therefore, the fundamental ingredient in our model is a set $S$ of abstract 
{\em states}, where each state serves as a possible description of the 
subject.  In general, we view the set of states
as enormous, since the states need to be able to recognize
fine-grained distinctions between subjects: if two subjects
differ in a way that might be relevant to some form of evaluation,
this should imply that these two subjects reside at different states.
For example, if one is evaluating a student's ability to multiply
numbers, then the state should be expressive enough to describe the
student's aptitude at multiplication and how they came to acquire this
aptitude. Similarly, if one is evaluating an athlete via a 100-meter
sprint, then the state should describe the athlete's sprinting
abilities, including information about their training up to this
point, as well as the conditions under which the race is run. The fact that states are expressive enough to capture
fine-grained differences between subjects means that for any
particular subject, it will not in general be possible to learn
their precise state from any limited amount of interaction with them.

Any property can be described by the set of states at which it holds,
and our discussion thus far has implicitly been concerned with
four properties that can in general all be different from one
another:
\begin{itemize}
\item Society's initial property of interest---the concept with
which we began---can be viewed as a set of states $I_s \subseteq S$.
\item The evaluator might have motivations that are
distinct from simply assessing the property $I_s$; thus, we assume
that the evaluator is interested in identifying subjects who belong
to some possibly different set of states $I_e \subseteq S$.
\item Since the state set is enormous, and states might differ in
hard-to-discern ways, it is generally not possible to perfectly evaluate 
a particular property; as a result, the set of states that
correspond to passing the evaluator's test is a set $P \subseteq S$ that
might be different from both $I_s$ and $I_e$.
\item Finally, the subject begins at some initial state $s_0 \in S$
and has a budget of effort that they can spend to move to a new state,
$s_1 \in S$, with the goal of reaching a state that passes the test.
Let $R \subseteq S$ be the set of all states that the subject can 
reach using this budget of effort; thus, it is possible for the subject
to pass the test if and only if there is a state in $R \cap P$ --- both
reachable and among the passing states.
\end{itemize}

\noindent
Now, for any collection of subsets 
$C_1, C_2, \ldots, C_k \subseteq S$, let's say
that two states $s$ and $s'$ are {\em indistinguishable} with respect to
$C_1, C_2, \ldots, C_k$ if for each $C_j$, 
the state $s$ belongs to $C_j$ if and only if the state $s'$ does.
Notice that indistinguishability is an equivalence relation, and
so it divides the state space into equivalence classes.
Since a given state can either belong or not belong to each of
$C_1, C_2, \ldots, C_k$, there are $2^k$ possible equivalence classes,
though some of them may be empty.

\begin{figure*}
    \centering    \includegraphics[width=.7\linewidth]{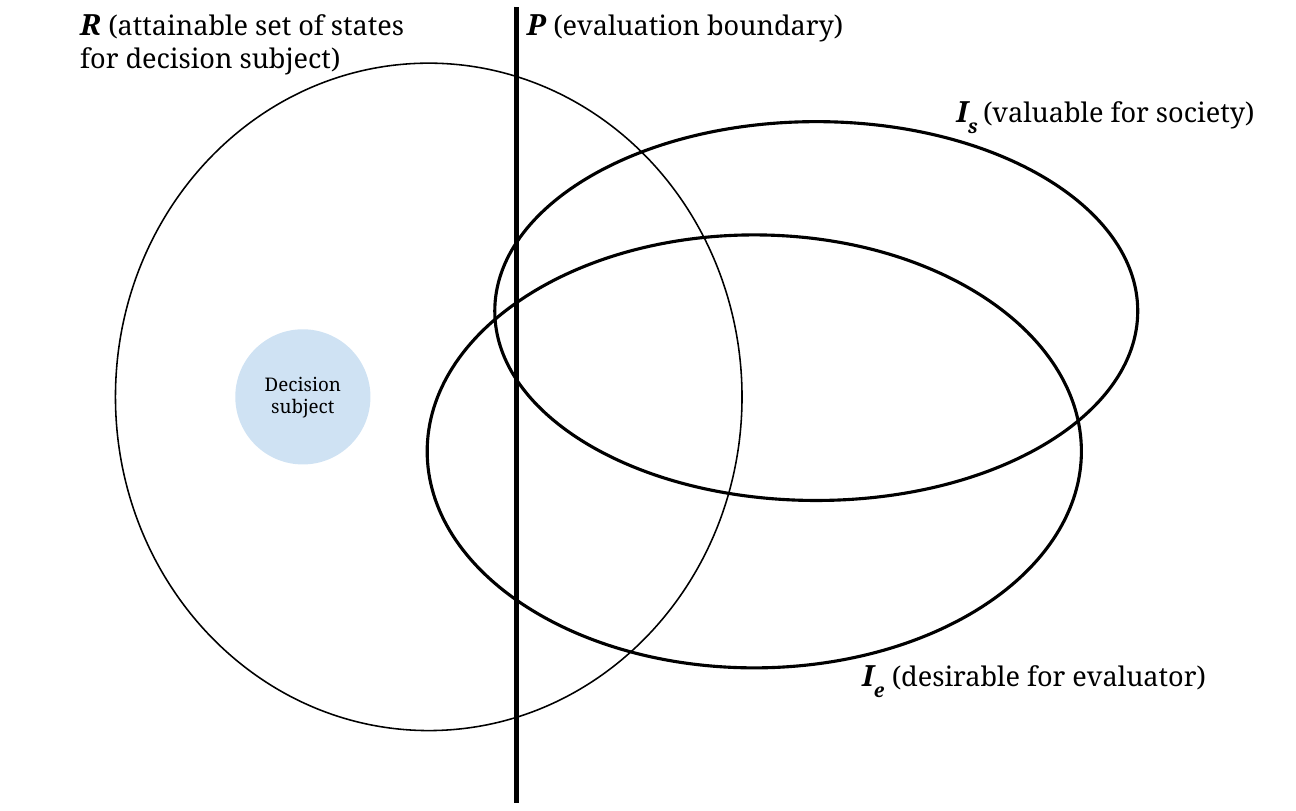}
    \caption{Diagram representing relevant states in the three-party model of evaluation. Each position for the decision subject represents a state which may or may not be described by any of the following four properties: Attainable for the decision subject ($R$), valuable for society ($I_s$), desirable for the evaluator ($I_e$), and passing the evaluation ($P$). Not all Boolean combinations are depicted.}
    \label{fig:my_label}
\end{figure*}

Using the four subsets of the state space we have defined---$I_s$, $I_e$, $P$, and $R$---the different scenarios of interest to us can be categorized
by whether a given state belongs (or does not belong) to each of
$I_s$, $I_e$, $P$, and $R$;
that is, there is a different scenario for each of the
$2^4 = 16$ equivalence classes
of indistinguishable states with respect to $I_s$, $I_e$, $P$, and $R$.
These categorizations are illustrated in Figure \ref{fig:my_label}.
As discussed above, a state's membership in a given equivalence class
does not convey normative information on its own, but this decomposition
into equivalence classes provides a starting point for ethical
analysis by systematically mapping the scenarios that can arise
according to these four underlying dimensions.

\section{Extended examples}\label{sec:examples}

In order to make the utility of our three-player model more concrete and to demonstrate how the interests of subjects, evaluators, and society can be variably aligned or misaligned, we discuss three example cases: hiring, grade inflation, and sports.

\subsection{Hiring} 
\label{subsec:golf-hiring}
In hiring, a variety of evaluations may be employed to assess candidates’ fitness for a position. We can think of hiring as a multi-stage process in which an initial pool of candidates is winnowed down into progressively smaller sets; evaluations are conducted at each stage in order to select the candidates who will progress through the pipeline \cite{bogen2018help}. Recently, a good deal of research has focused in particular on initial screening steps in the hiring process. These could include algorithmic tools to analyze resumes; personality quizzes, games, and analysis of video interviews to predict a candidate’s likelihood of job success; or the degree to which candidates exhibit certain qualities, like resourcefulness or grit \cite{raghavan2020mitigating}. Much attention has been paid to the fairness and diversity implications of these tools, many of which are poorly validated. Beyond initial screening stages, evaluations commonly involve candidate interviews and other face-to-face activities with hiring managers or prospective colleagues, ostensibly designed to assess aptitude and ability to respond to questions or solve problems related to the work task, or to gauge a candidate’s collegiality and ``fit’’ with workplace norms and culture. 

A two-player model of strategic behavior might direct attention to how candidates seek to present themselves to the hiring firm in order to signal their aptitude or qualifications for the job, given the types of evaluations to which they expect to be subjected by the hiring firm. They might do so, for instance, by tweaking their resumes---say, by describing accomplishments using terms likely to be viewed favorably by an algorithm, accumulating ``fluff'' credentials, or even providing false or exaggerated information about previous accomplishments. Or they might seek to signal cultural ``fit'' at an in-person interview via clothing choices, comportment, or chosen topics of conversation.

Viewing the design of the evaluation, itself, as a strategic activity---and concomitantly, the evaluator (the hiring firm), itself, as a self-interested actor---foregrounds new questions. In a two-player model, a firm’s evaluation is implicitly treated as representative of broader societal interests. But in a three-player model, the firm's own goals in conducting the evaluation may diverge from goals that are aligned with societal welfare or normative values. Social scientific research suggests that this divergence is not uncommon. Firms are known to implement evaluations in order to hire candidates that ``fit,’’ a notoriously slippery concept, which often manifests as demographic homogeneity with hiring managers and other current employees. Cultural ``fit'' with a firm may thus diverge from societal ideals of merit, fair treatment, and nondiscrimination in employment. Ray’s \cite{ray2019theory} theory of racialized organizations, for example, describes mechanisms through which seemingly neutral hiring and credentialing processes in firms can be racially exclusionary despite legal antidiscrimination mandates. Rivera’s ethnographic research \cite{rivera2016pedigree} on hiring at elite firms shows that applicants from well-resourced backgrounds do better across all stages of the hiring process due to a combination of economic advantages, social connections, and cultural resources that signal their social position to gatekeepers (i.e., hiring managers) (see also \citet{bourdieu1987distinction}).

Consider the following hypothetical example. In some law schools, student services staff sponsor golf instruction for law students who are unfamiliar with the game. Though golf lessons may seem orthogonal to the ostensible substantive goals of legal education, the lessons are designed to equip students with the \textit{cultural} toolkit for job success. Golf is, traditionally, a sport strongly associated with economic privilege, and one historically off-limits to women and non-white players; as such, privileged white men are more likely to know how to play golf. Given the reality that many elite law firms are also disproportionately composed of privileged white men, and that those firms may seek to hire ``the kinds of people’’ who know how to play golf (i.e., golf-playing is an indirect proxy for whiteness, maleness, and socioeconomic privilege), golf instruction in law schools can be understood as a means of trying to assist law students in signaling cultural fit. Indeed, law school golf programs are often explicitly directed toward women, and place emphasis on basic golf etiquette and literacy as well as skills. (For instance, participants in a golf program for women students at Arizona State University’s Sandra Day O’Connor College of Law acknowledged that they were learning to play because ``golf is an access issue’’ and that they ``didn’t want to be left out’’ of the networking opportunities that knowing how to play golf could yield \cite{anderson2022asu}).

Understood through the lens of our model, we can envision a law school graduate, seeking a job at an elite firm, who is competent in the practice of law (thereby belonging to a state in $I_s$, society’s property of interest). Imagine that our hypothetical job seeker does not come from an elite socioeconomic background, and has not played golf before. By taking golf lessons and developing a cultural facility with golf, she is able to---and does---pass the evaluator’s test (that is, successfully interview for the job) in a hiring process; familiarity with golf has enabled her to reach a state that belongs to both $P$ and $R$. (Even if the hiring process doesn’t include an explicit golfing component, we could imagine several ways in which this cultural knowledge might surface in an interview---for example, through discussion of a recent PGA tournament, or conversation with the hiring manager about local courses.) However, if we imagine that for the elite law firm, golf skills are valued because of their traditional correlation with a particular class background, and have served as a ``cultural fit’’ proxy for reproducing the current demographics of the firm among new hires, then our job seeker has \textit{not} attained the evaluator’s property of interest, $I_e$; that property is out of alignment with the others in our model.

As we’ve described, by considering alignments and misalignments among these states, our three-player model provides a mechanism for directing our attention to ethical implications of strategic behavior with more nuance. In a two-player model, we might simply view our job-seeker’s golf lessons as strategic behavior against the evaluator’s goals, and might seek ways to limit or discount its influence on the hiring process. The three-player model shows how the evaluator’s interest $I_e$ is itself out of step with both society’s interest $I_s$ and the interest of the job-seeker, who wishes to reach a state in $P$. Instead, golf lessons may be recast as an effort to push back against an unjust exclusionary criterion, which serves to align the subject's strategic behavior with societal objectives. As such, it may be judged as ethically acceptable.

Further, this example demonstrates an additional benefit of the three-player model. Much contemporary scholarship on fairness in hiring processes focuses exclusively on screening stages, when algorithmic tools are used to winnow down a set of candidates to a set to be ``called back’’ for an interview. (Most social science audit studies of these processes also focus exclusively on these early hiring stages, as demonstrated by Quillian et al. \cite{quillian2020evidence}, for both pragmatic reasons and based on research ethics considerations.) But a good deal of biased and exclusionary hiring practice---that is, misalignment of an evaluator's objective and societal objectives---is likely to occur in \textit{interview} stages, which are often excluded from scrutiny by researchers studying the ethical dimensions of AI-driven tools or the fairness of hiring processes \cite{vecchione2021algorithmic,narayanan2022limits}. A broader conceptualization of strategic behavior offers a more inclusive ``end-to-end’’ view of the hiring process and the ethical dimensions thereof.

\subsection{Grade Inflation}

\textit{Grade inflation} is a phenomenon in which the grades students are assigned for coursework tend to ``inflate’’ (i.e., increase) over time. Empirical data demonstrate the existence of the phenomenon across colleges and universities: in one study of 200 U.S. schools \cite{rojstaczer2012ordinary}, ``A'' grades comprised 43 percent of all letter grades issued in 2009, as compared to only 15 percent in 1960. Grade inflation is particularly pronounced among private colleges and universities, even controlling for student selectivity \cite{rojstaczer2012ordinary}.

How might we understand grade inflation through the lens of strategic evaluation? We can conceive of the evaluation as the issuance of a grade (or a set of grades) to a student based on course performance.
If we think of $P$ as representing the set of states resulting in a high grade, then the student has an interest in finding a state in $P$ and $R$: a reachable state in which they receive a high grade.
The student presumably benefits from high grades (e.g., as a credential for a future job search). The educational institution, as evaluator, also has interests in issuing high grades: they have a reputational interest in ensuring that graduates are successful on the job market, and high grade point averages can help to set students up for such success. When grade inflation is particularly widespread, schools may find it difficult to ``deflate’’ due to concern about harming students’ career opportunities or becoming less competitive in recruiting new students \cite{kohli2014princeton}. Schools may also be interested in keeping students satisfied via high grades for purposes of maintaining positive alumni relations, which bear reputational and economic dividends (e.g., they might result in greater donations to the school in the future).
Accordingly, the institution has incentives to make sure that the set of states $I_e$  it approves of contain many states in $P$ and $R$: reachable states conferring high grades.

Instructors, whom we can think of as acting as agents of the educational institution, have their own set of interests, which might support grade inflation. Instructors may enjoy better interpersonal relationships with students when they assign them high grades, and empirical evidence suggests that instructors who issue high grades receive better evaluations from students, which may factor into faculty’s own tenure and promotion evaluations \cite{lanning1995grade}. (We could also, of course, conceive of instructors and educational institutions as separate ``players’’ in our model that are potentially \textit{misaligned} in their interests regarding grading; we collapse them here for the sake of simplicity in illustration, since for purposes of our discussion they both have incentives for $I_e$ to contain many states in $P$ and $R$.)

Thus far, then, both the subject and the evaluator may be aligned in their interests, supporting an inflated grade regime. But society's interest---represented as $I_s$---may be misaligned.
Indeed, we can think of the ``inflation'' of the grades as an enlargement of the set $I_e$ relative to the set $I_s$: the institution is willing to view many more states as deserving of high grades, whereas society might want $I_s$ to be a smaller set that has fewer states in $R$ (corresponding to fewer states achievable by students).

There are many potential arguments for why societal interests might be poorly served by inflated grades. Grade inflation might be problematic to the extent that grades are useful tools for distinguishing among the performance of different students; if everyone gets an ``A,'' it's not as clear which students truly excelled \cite{kohn2019why}. Similarly, some argue that grade inflation may diminish students' motivation to excel in coursework, because attaining a top grade takes relatively less work, thereby reducing students' capacity to reach their full learning potential. The ethical and social implications of grade inflation are strongly debated, both in the academic literature on the topic \cite{crumbley2010ethical, finefter2015wrong, lanning1995grade, rojstaczer2012ordinary} and within institutions facing public pressures to rein in inflation. In situations like these, interests $P$ and abilities $R$ of the subject are aligned with the interests of the evaluator $I_e$, but misaligned with societal interest $I_s$, as illustrated by our three-party model.

\subsection{Sports} 
The sports world is also a useful site for illuminating these dynamics,
owing to its intentionally competitive design. In sports, acceptable
means of reaching a particular objective (scoring, or winning a race) are generally made explicit via a set of detailed rules
promulgated by the sport’s association or governing body, and
subsequently enforced by referees or other officials.

Sports, therefore, are a natural setting for studying strategic behavior.
A traditional analytic structure would posit that 
the organizer of the sporting event (the evaluator) 
creates a set of rules designed to measure athletes' (subjects) abilities, while  athletes look for ways to gain an ``edge'' within
the constraints of the rules.
In particular, to prepare for an event, athletes train to
improve speed, strength, and agility; they strategize about and
prepare for likely competitive scenarios; they gather information about the
strengths and weaknesses of the competition. A number of sports
scandals and armchair debates involve the normative bounds of these
behaviors, when such strategic efforts cross a line from gamesmanship
into territory disallowed by the organizers---including
doping, illegal sign-stealing, and other forms of (what is commonly
perceived as) ``cheating.’’

Consider, for example, a championship-level track and field meet.
We often think of such events as having some of the most straightforward
specifications---to run a certain distance as fast as possible,
or to jump as far as possible---but of course they are also
controlled by rules concerning allowable equipment, racing conditions,
and substances (e.g. drugs) that an athlete is or isn't allowed to ingest
while training or competing.
We intuitively think of the ``organizers'' of the event 
as the enforcers of the rules, where the organizers represent some 
amalgam of the local operations of the meet and the international
governing bodies for track and field.
The two-party analysis of this setting would take this collection
of organizers to be the evaluator, formulating and enforcing rules
that apply to the athlete as subject.

As with the other domains we've considered, this two-party
interaction between the evaluator and the subject misses a number of 
the central issues that arise in the process of governing a sport
like track and field.
A salient example is the design of the track itself: at the 2021 Tokyo Olympics, a great deal of technology and money
went into the creation of a ``springy'' track surface (i.e.  rubber granules
for better shock absorption) to enable the runners
to increase their speed and increase their chances of breaking records
\cite{nyt-olympic-track}.
This type of technology could be pushed much further than it was
at the Olympic Games; what determined the limit of the track's 
springiness was not technology, but a sense that going too far
would risk the athletes' safety, and cross a notional line that
separates the act of running on a track from the act of running
across a 400-meter trampoline.
This notional line was therefore enforced by material-science
specifications for allowable track surfaces defined by
World Athletics, which governs track and field events
\cite{world-athletics-track-surface}.

In one sense, 
this example reveals a familiar kind of strategic interaction: the
hosts of the track meet spend money to commission a track whose
surface pushes up against the allowable specifications, and 
the governing body enforces rules designed to preserve the underlying
intent of the activity.
But in another sense, this strategic tension is happening {\em within}
the set of parties that a simpler analysis might have grouped 
together as a single ``evaluator.''
This is precisely the richer view that a multi-party analysis makes
possible: 
athletes would like to win races, and they work strategically within the
rules enforced by the event organizer and the governing body
to achieve this; 
event organizers would like to host track meets where world records
are set, and they work strategically within the rules enforced by
the governing body to achieve this.
The 2021 Olympics is far from the only recent high-profile example
of these issues in track and field;
another recent instance is the attempt (with help from Nike as part of its {\em Breaking2} campaign)
to create approved conditions under which it would be possible
for an athlete to run a marathon in under two hours
\cite{wired-two-hour-marathon}.
It is worth noting that once we move from a two-party view
to a multi-party view, there is no reason we need to stop at
three parties;
for example, even the governing body of track and field is motivated to create conditions in which dramatic events happen 
in their sport in order to attract publicity and attention. In doing so,
they operate strategically: for example, choosing how to
set standards within informal constraints set by further parties, including the opinion of the public and the sports media
about what constitutes a reasonable format for
the event. Similar considerations arise in many other sports.  

A point worth highlighting is the contrast between technical restrictions on
an allowable track surface and technical restrictions on  allowable equipment,
such as running shoes (which, like the track, are also made of rubber and
designed to be springy). Though they appear similar, a key contrast is that strategic innovations in equipment
are made by parties who are helping athletes; in contrast, strategic innovations in track surfaces are made by parties whom we typically think of as maintaining the integrity of the event---but whom, according to our model, we can also view as strategic actors motivated in part by their own aims.

Since there are multiple scenarios in these settings that differ
in subtle ways, our formalism in terms of subsets $P$, $R$, $I_e$, and $I_s$
can help clarify the distinctions among them.
In applying the formalism to our examples here, 
we think of the underlying states as representing not only qualities about a competitor, but also different
track meet scenarios and their outcomes. 
We focus on some definition of success ---
such as whether a particular world record has been broken.
$P$ is then the set of states where this success outcome occurs;
$R$ is the set of states achievable by the athlete;
and $I_e$ and $I_s$ are the states that are acceptable to
the local organizer of the event and the global governing body, respectively.
In this way, we can distinguish among the interpretation
of states based on whether or not they belong to each of the four sets:
\begin{itemize}
\item We start with the most straightforward case,
in which an athlete breaks a record under conditions that are
acceptable to both the event organizer and the governing body.
This is simply a state in all four of $P$, $R$, $I_e$, and $I_s$.
\item Now consider the following hypothetical scenario, inspired by our discussion: an event organizer commissions a highly springy track,
resulting in a new world record; but the track is later found to violate the allowable material-science specification for track surfaces.
This would correspond to a state in $P$, $R$, and $I_e$, but not $I_s$.
\item Thus-far unsuccessful efforts to produce a marathon time under
two hours using fully approved marathon conditions correspond to a 
different type of state: the organizers and the governing body work
together to formulate a state in all three of the sets $P$, $I_e$, 
and $I_s$ --- i.e., an approved outcome that breaks two hours ---
but because human runners are not able to attain this state, it is not
in the set $R$ of states reachable by the subject.
We can think of it as an open question whether --- for this activity,
with $P$ corresponding
to marathon times under two hours --- there in fact exists a state
in all four of the sets $P$, $R$, $I_e$, and $I_s$ \cite{joyner1991modeling}.
\item There are other scenarios that follow almost mechanically;
for example, if an athlete breaks a world record, is subsequently 
disqualified by the local organizers, but has their time reinstated after a 
successful appeal to the governing body, this corresponds to a state
in $P$, $R$, and $I_s$, but not $I_e$.
\end{itemize}

\section{A Mechanical Understanding}
\label{sec:mechanical}

In the previous section, we discussed a number of evaluation scenarios in which social dynamics lead to strategies that may or may not serve the interests of evaluators, decision subjects, and society. We now argue that our model (described in Section \ref{subsec:model}) doesn't just capture these individual scenarios, but extends to a wide range of candidate and evaluator behaviors in various domains. Where the previous parts of this section focused mainly on identifying particularly evocative examples of behaviors and strategies, we suggest here that the model can also be useful in an \textit{enumerative} role: By varying its parameters, our model is able to portray all the possible stories we set out to describe. Given its structure as a three-party game, we can use the model to enumerate the scope of possible scenarios -- both mundane and nuanced -- where the interests of a subject, an evaluator, and society either align or diverge.

\begin{table*}
    \centering
    \begin{tabular}{|c|c|c|c|}
        \hline
        \makecell{Would passing serve\\ societal values?\\$s_1 \in I_s$} & \makecell{Would passing serve\\ evaluator's interests?\\$s_1 \in I_e$} & \makecell{Passes?\\$s_1 \in P$}  & \makecell{\textbf{Example states and scenarios}}  \\
       \hline
        \cellcolor{green!40}Y & \cellcolor{green!40}Y & \cellcolor{green!40}Y & \makecell{Candidate has a strong record and grew up playing golf.\\ She passes the interview.
        }\\
        \hline
        \cellcolor{green!40}Y & \cellcolor{green!40}Y & \cellcolor{red!30}N & \makecell{Candidate has a strong record and grew up playing golf.\\ She fails because she got sick on the interview day.
        }\\
        \hline
        \cellcolor{green!40}Y & \cellcolor{red!30}N & \cellcolor{green!40}Y & \makecell{Candidate has a strong record and hadn't played golf.\\ She passes thanks to a golf program in law school. 
        } \\
        \hline
        \cellcolor{green!40}Y & \cellcolor{red!30}N  & \cellcolor{red!30}N & \makecell{Candidate has a strong record and hadn't played golf.\\ She fails because she had few opportunities to network.
        }\\
        \hline
         \cellcolor{red!30}N & \cellcolor{green!40}Y  & \cellcolor{green!40}Y & \makecell{Candidate has a weak record and grew up playing golf.\\ She passes the interview after networking over golf.  
         }\\
        \hline
        \cellcolor{red!30}N & \cellcolor{green!40}Y  & \cellcolor{red!30}N & \makecell{Candidate has a weak record and grew up playing golf.\\ She fails, despite networking, due to a positive drug test.
        }\\
        \hline
        \cellcolor{red!30}N & \cellcolor{red!30}N  & \cellcolor{green!40}Y & \makecell{Candidate has a weak record and hadn't played golf.\\ She passes after lying about her experience.
        }\\
        \hline
        \cellcolor{red!30}N & \cellcolor{red!30}N  & \cellcolor{red!30}N & \makecell{Candidate has a weak record and hadn't played golf.\\ She fails the interview.
        }\\
        \hline
    \end{tabular}
    \caption{A mechanistic example of different states. In this stylized story, a hiring decision is heavily dependent on an interview process in which playing golf --- and an upbringing that involved time spent around golf courses --- sometimes plays a significant role. We assume that this hypothetical evaluator is interested in golf-playing candidates in order to highlight the ways that an evaluator's interests can diverge from societal goals and values. The example scenarios are individual instances within a broader set of states.}
    \label{tab:mechanistic}
\end{table*}

Consider a particular domain where the subject of an evaluation (e.g., a job candidate or athletic competitor) is described by some state at the time she is evaluated. Recall we defined this state as $s_1 \in S$, the state a subject might occupy after exerting strategic effort. There are three qualities of this state which, we believe, capture much of the important social context for reaching descriptive or ethical conclusions about the evaluation. The first important quality, and perhaps the most straightforward, is whether the state `passes' the evaluation (or otherwise performs favorably). This is true if $s_1 \in P$, where the group of passing states $P$ is defined by criteria that are decided on, strategically, by the evaluator. The second important quality of state $s_1$ is whether it genuinely represents an example of the quality that is desired by broader social interests. Does the subject actually excel at the activity that is supposedly being tested for? If so, we would say $s_1 \in I_s$. Finally, the third important quality is whether the candidate's state $s_1$ serves the strategic interests of the evaluator. If passing the evaluation subject would be strategically beneficial for the evaluator, then $s_1 \in I_e$. All together, these qualities can help us characterize and make sense of the outcome of an evaluation. 

For ease of exposition in this section, we leave out the (fourth) question of whether $s_1$ is a feasible state for the decision subject to reach (i.e. whether $s_1 \in R$). It would not be difficult to extend our discussion to include a distinction between whether $s_1$ belongs to $R$ or not, but for now we focus our analysis on states in $S$ without including this distinction.

We therefore have a taxonomy of states based on distinguishing whether a state $s_1$ belongs to $P$, whether it belongs to $I_s$, and whether it belongs to $I_e$.
For each combination of these three different qualities, we provide an example scenario that intuitively satisfies the given combination.
To show how all the possible scenarios can arise in a single unified setting, we situate all of them in a stylized story of a law student applying for a job at a prestigious law firm.
For pedagogical purposes, we imagine that the law firm has an overtly (and in our telling, somewhat cartoonishly) biased hiring process that grows out of the discussion in Section \ref{subsec:golf-hiring}; specifically, the firm seeks candidates who grew up in affluent circumstances, and they attempt to discern this by inviting job applicants to play a round of golf during their on-site interview visit.
Our formulation is therefore deliberately extreme so that it can make clear the distinctions among different scenarios; in more nuanced situations, we would have the same formal structure but potentially a more challenging interpretive task in distinguishing among different scenarios.

Because our model contains membership in the sets $P$, $I_s$, and $I_e$ as yes/no predicates, we do not need to be inventive to list a range of different scenarios in which a prospective lawyer applies to such a firm; rather, we can literally build a table that mechanically enumerates all possible outcomes for these predicates.
We do this in Table \ref{tab:mechanistic}.
For example, consider the hypothetical scenario where a fantastic and resourceful law student from a low-income background takes advantage of golf lessons offered through a university. If she receives a job offer in part because she was able to network with a partner over golf, her scenario represents a particular entry in our table. Since she is an otherwise fantastic lawyer, her candidacy for the job serves society's interests $s_1 \in I_s$. However, the partner's preference for networking over golf suggests a latent prejudice, in which the candidate's background does not serve the internal interests and preferences of the firm, $s_1\notin I_e$. However, the candidate does pass, $s_1 \in P$. This scenario is depicted in the row of Table \ref{tab:mechanistic} corresponding $s \in P$, $s \in I_s$, $s \not\in I_e$. Now imagine a similar scenario, equivalent in every way, except that the candidate does not network over golf, and instead passes the evaluation because of her strong track record and depth of legal knowledge. This more straightforward case -- where the candidate does not face prejudice in the evaluative process, and the firm simply hires somebody based on their strong record -- corresponds to row 1 in Table \ref{tab:mechanistic}. 

There are a few possible ways to interpret the taxonomy. Notice first that disparities between the evaluator's interests and society's (i.e. places where a state is in one of $I_e$ or $I_s$ but not the other) often suggest a place where the evaluator is applying a bias that is not societally beneficial.
Next, we observe that absent any regulatory intervention on society's behalf, disparities between the evaluator's interests and the outcome of the test represent \textit{noise} or \textit{error} in the evaluation. This is because if an evaluator has full reign over the criteria and standards composing the assessment, and still a candidate faces an outcome that is out-of-step with the evaluator's interests, then this simply suggests a noisy, or error-prone evaluation. A final observation is that, generally, a good evaluation is one where all probable and feasible states $s_1$ pass ($s_1 \in P$) if and only if they serve societal interests $s_1 \in I_s$. In other words, a desirable evaluation is one where the values of these two predicates should match.

\section{Taking an ethical perspective} 
\label{sec:ethical}

The framework developed so far, and expanded through the examples in the previous sections, provides several useful perspectives on the process of quantitative evaluation. First, it allows us to appreciate that evaluators can act strategically in their own interests: in other words, that gaming and strategic behavior are not only carried out by the subject of the evaluation, but by the parties designing the evaluation as well. In this way, it helps to recast normative judgments about a subject's behavior, interpreting deviations not as much from a fixed point but, instead, in light of the evaluator's own aims, which may themselves warrant scrutiny. 

The framework sheds light on the disparities that may arise between the interests of the evaluator and societal interests (i.e. social welfare) more broadly, whether expressed through collections of norms or through explicit regulation. By making these disparities a central focus of our framework, we can distinguish between cases in which the evaluator makes these disparities explicit as part of the evaluation itself, and cases in which these disparities remain covert and require additional scrutiny to unearth. This distinction is crucial for work that brings AI and machine learning to bear on quantitative evaluation: current work in this space has tended to shine a spotlight on the explicit, quantitative components of decision-making processes, giving less attention to the parts of the decision-making pipeline where an evaluator's aims might be unstated, under-specified, and difficult to discern. Our model, in which the evaluator is cast as a strategic actor, opens up the possibility of turning new modeling attention to these more implicit (and possibly covert) parts of the decision process (see \citet{barabas2020studying}).

Our model therefore also broadens the notion of governance and regulation for an algorithmic system that makes and enforces rules. Rather than conceiving of rule-making and rule-enforcement processes as undertaken by a single monolithic entity, as much of the literature on strategic behavior has tended to do, it becomes more tractable to analyze the internal tensions that exist within this process, between multiple rule-making parties with potentially divergent interests. Attempts to address strategic behavior in algorithmic systems, via either technical or policy mechanisms, are well-served by recognizing these complexities in real-world evaluations.

Finally, our perspective has important implications for the moral scrutiny to which strategic behaviors are often subjected. A student found cheating on a math test, or an applicant embellishing a resume, might reasonably draw moral disapproval for those actions, and on its grounds warrant rejection, penalty or down-weighting. The language we use to describe such actions, e.g. ``cheating,'' ``gaming the system,'' or ``honest effort,'' may convey ethical meaning, prejudging a case even when the underlying behaviors might be ambiguous. A distinctive virtue of our framing is that by conceiving of evaluators as potentially strategic actors, too, it allows the same moral scrutiny to be directed to them and not merely to those being evaluated.

\textit{Judgments about subjects depend on judgments about evaluators.} Normally, the outcome or score yielded by an evaluation mechanism is taken as legitimate grounds for choosing one candidate over another, or declaring one competitor victorious over another. If an evaluation mechanism is considered sound, that is, is considered to be an effective or reliable measure of a target quality, candidates who strategically alter their features to ``beat'' the mechanism in ways unforeseen, or unaccounted for by the evaluator are, in the first place, presumed to be behaving unethically. Some have pointed to differences among such workarounds that justify classifying them in different ways. Miller \cite{miller2020strategic}, for example, classifies actions taken by a decision subject which are causally linked to the intended outcome as ``improvement,'' otherwise, as ``gaming,'' where the latter suggests unjustified success on a given performance metric.
Greater subtlety in assessing subjects' behaviors in ethical terms seems important, but even here the focus of moral scrutiny has not shifted away from the subject of evaluation. Although the \textit{quality} of a given mechanism may be called into question for failing effectively to measure a target, the target of a measurement itself, typically, is taken as given, a fixed point, which is assumed to align with societal values \textit{a priori}. 

Our framing treats target states as variable. In so doing it insists that the goals and methods of an evaluator are relevant factors in assessing the moral standing of decision subjects' strategic behaviors. Examples of morally questionable measurements include discriminatory, elitist, or overly demanding hiring procedures; even including some which purport to serve values like diversity, meritocracy and fairness. Tests can play insidious gate-keeping roles, and sports standards can be mired in corruption. Acknowledging that evaluators' strategies potentially diverge from societal mandates means that it may be unjust to pin responsibility for strategic behaviors solely on a decision subject. These dynamics suggest a need for a more nuanced, contextual assessment of the behaviors of decision subjects that takes into account the legitimacy of an evaluator's aims and methods. As the vast literature on lying suggests, absolutism aside, an act of lying may range in moral standing depending on morally relevant contextual considerations \cite{bok1999lying}.

We can apply these arguments to our earlier example of law schools offering golf instruction.
Such a behavior is a strategic response to an existing norm among law firms---namely, that a significant amount of communication and fraternization occurs over golf. Such a norm might introduce bias into hiring and promotion processes, because people comfortable and experienced on golf courses tend to be whiter and wealthier. As such, spending time in law school learning golf, though not traditionally thought of as causally linked to successful law practice, might be justifiable given the contextual norms and evaluation criteria. Passing negative moral judgments (or, scoffing) at people learning to play golf misses the broader social forces inducing such behaviors, and can work to further exclude people not socially positioned to learn golf at a young age.

\textit{Evaluators can behave deceptively.} As we noted in Section \ref{overview}, there are sources of slippage between performance on a test and a true property of societal interest. These can be described succinctly by the differences between $I_s$, $I_e$ and $P$.

When an evaluator's goal states $I_e$ are transparent and accessible, rifts between $I_e$ and $I_s$ may be more easily identified. Observed discrepancies might require new forms of oversight and standards-setting so that evaluations do not skew towards states favorable to the evaluator that diverge from societal aims and values. These discrepancies show up in our example of Nike's {\em Breaking2} campaign, in which the company sponsored a race that aimed to enable athletes, sporting Nike sneakers, to complete a marathon in under two hours. This goal diverges from traditional marathon goals which aim towards consistency among races. Nike's highly publicized  goal drew attention, by design, to the ways that it set up the race conditions to help its runners, e.g. providing pacers and positioning them to reduce wind resistance. Even though Nike's sponsored runners outpaced the standing marathon world record, the explicit highlighting of changes to these conditions helps make clear the ways in which the improved time would not have met the governing body's requirements for an official world record.  

When an evaluator's goals are unclear, misleading or deceptive, however, it can be difficult to draw normative conclusions or assign responsibility. Differences between passing states $P$ and societal goals $I_s$ might arise because of benign and necessary practical limitations, such as measurement error, or they might arise because of blameworthy and pernicious strategic behaviors on the part of the evaluator. Undoubtedly, it is not always clear whether an evaluator is acting perniciously or in good faith. For example, academic departments, lacking diversity, could point to larger systemic issues that create barriers for under-represented minorities and claim that their interests and intentions are aligned with societal efforts to increase diversity. As non-diverse hiring persists, however, we may have cause to question whether these arguments are given in good faith.

When an institution claims to have noble goals but outcomes diverge from $I_s$, a possible explanation is that it is behaving strategically according to \textit{covert} interests in $I_e$. By citing universal and unavoidable issues around measurement instrumentation, evaluators may be afforded some \textit{wiggle room} to dodge normative scrutiny even when they are acting in ways that are counter to society's interests.\footnote{Furthering covert goals of the sort we describe might constitute manipulation, defined as \textit{hidden influence} \cite{susser2019online,susser2019technology}.} Examples include college admissions, where the underlying concept of societal interest --- college aptitude --- is fundamentally contested. In light of the broad disagreement over appropriate norms and standards, colleges employ concepts like `holistic review' which are notably opaque. These strategies successfully avoid the fundamentally value-laden questions about what college aptitude is. They also afford some potentially self-dealing behavior, like earmarking applications from friends of trustees. 

The main point of this section is to emphasize that strategic gaming might be tolerated or possibly even encouraged when evaluators, through their evaluation mechanisms, reward capacities that are not in alignment with societal ends and values. Furthermore, strategic gaming may be ethically justifiable to achieve a favorable outcome in competitions where the rules have not been designed in good faith. As such, 
it is important to remain astute to efforts by evaluators to obfuscate evaluation mechanisms that are misaligned with societal values, which reduce the capacity to identify relevant excusing conditions and also the capacity to reliably assign moral responsibility within such systems, overall.

\section{Further related work}
\label{sec:related}

Here, we connect our work to the formidable literatures on both strategic algorithmic systems and theories of evaluation. We do not aim to provide an exhaustive review, but rather to situate our contribution and highlight relevant work. Our takeaway is that evaluation scenarios necessarily invoke \textit{societal values}. These encoded values and commitments clarify the moral standing of strategic behaviors.

\subsection{Strategic Behavior}

A growing body of theoretical and applied work in machine learning focuses on dealing with strategic behavior and distribution shifts in response to algorithmic decisions. By this view, algorithms and metrics influence decisions that have an effect on the people and systems being measured, who therefore behave strategically to attain a desired outcome.  

 In literature on \textit{strategic classification} \cite{hardt2016strategic}, this dynamic is expressed using a simple game: An evaluator (player 1) first announces an evaluation scheme, and a decision subject (player 2) responds by investing some effort to alter his features and potentially change his classification outcome. The evaluator’s goal is strong performance on a metric like classification accuracy in light of potential distribution shifts caused by strategic responses, which impede the evaluator’s ability to observe the decision subject’s ``true'' underlying labels. A variety of related models have been put forward for achieving a similar goal in other statistical settings, like regression \cite{chen2018strategyproof,shavit2020causal,harris2021stateful}, ranking \cite{liu2022strategic}, and in repeated games \cite{hu2018short,zrnic2021leads,holmstrom1999managerial}.

\textit{The influence of mechanism design.} Strategic models of ML tasks inherit a core assumption from behavioral economics and mechanism design: that social systems can be modeled as interactions among agents who behave according to rational self-interest. The tools of mechanism design have proven useful in designing systems with multiple agents to achieve certain desirable outcomes \cite{grossman1992analysis}. As a result, approaches drawing from the mechanism design literature tend to focus on a certain set of goals that the evaluator might have related to the integrity and effectiveness of the assessment---for example, preventing strategic behavior from occurring in the first place, or salvaging ``true'' signal from manipulated features.\footnote{It has been observed that applications of mechanism design can fall out of step with societal goals \cite{hitzig2020normative,viljoen2021design}. Meanwhile, attempts to align mechanism design with broader social interests are burgeoning. See, e.g., \citet{abebe2018mechanism,finocchiaro2021bridging}.
} These approaches typically conceive of the evaluator as solely interested in achieving a set of goals vis-a-vis the evaluated party (the decision subject).

\textit{Social costs, disparate effects.} In response, a chorus of literature invokes the ``social costs'' involved in algorithmic evaluation. These papers point out that assessments often involve powerful institutions setting the terms for distributing welfare and directing life outcomes---for example, through credit scoring, the provision of standardized tests in education, or the use of automated assessments in hiring. In defining societal considerations, these works tend to highlight the impact of an assessment on decision subjects. 
\citet{kleinberg2020classifiers} consider certain forms of strategic effort as utility-improving for both evaluator and decision subject. \citet{milli2019social} consider decision subject effort as a social cost that should be minimized. 
\citet{hu2019disparate} consider fairness in this context, finding that classification can exacerbate inequalities if decision subjects are afforded different budgets to strategically alter their features. By explicitly considering the impact of an evaluation on its subjects, these papers exemplify some of the social and ethical dimensions of evaluative settings where an evaluator's interests are misaligned with subjects'. 

Writing on the regulation of algorithmic decision-making and legal requirements aimed at transparency, \citet{cofone2019strategic} find that algorithmic decisions tend to involve strategic behavior both from decision-makers and subjects. Decision-makers, who often cite undesirable `gaming' from subjects as a reason to keep algorithms opaque, implicitly presume that that their interests align with society's. The paper makes the observation that the goals and behaviors of either player can be, plausibly, out of step with societal interests, so gaming may or may not be desirable. There is existing empirical work, too, on the contested and nuanced ethical boundaries of behaviors described by some as `gaming the system,' especially in the context of internet platforms, where content creators use search engine optimization \cite{ziewitz2019rethinking} and other practices geared towards courting viewers \cite{petre2019gaming}.

In our hiring example, the practice of golf lessons arises not as malicious or manipulative efforts among decision subjects, but as a response to a practice among the evaluating law firm(s). 
Our model attempts to describe the fact that \textit{both} the evaluator and the subject engage in strategic behaviors to achieve their own interests (which may, or may not, diverge from societal goals). 
Thus, we believe, a fundamental question that must be asked in evaluative contexts is: what are the appropriate societal goals underpinning an evaluation? Answering this question may enable society to institute mechanisms that make sure evaluations serve these goals, especially in cases when institutional interests diverge from society's. 
To that end, our work re-visits models of evaluation and draws conclusions not just about the appropriateness of responses, but about the appropriateness of evaluation measurements themselves. 

\subsection{Evaluation}

\textit{Evaluations} use measurements and observations to make a judgment of merit, worth or value \cite{scriven1991evaluation,scriven2007logic}. Although evaluations always involve empirical observation, not all forms of empirical observation constitute evaluation. Counting the number of yellow cars that pass on a highway is empirical measurement but is not an evaluation per se, because it does not help make a conclusion of merit, worth or significance.

Notice that many real-world settings with strategic behavior involve an evaluation. Grades measure educational aptitude. Sports measure athletic excellence. Job interviews measure skills, experience and fit. Settings involving high-stakes social decisions frequently draw from measures to assess constructs that are, at times, murky. Qualities like deservingness, or promise, or good business instincts -- these can be very difficult to ascertain. Evaluations may provide institutions with a seemingly less arbitrary way of making high-stakes decisions or allocating social welfare.

When an institution is tasked with conducting an evaluation, there is typically some societal \textit{interest}, or \textit{value}, that a set of people wish to measure. That value can be highly contested---as with intelligence---or comparatively less contested---as with sprinting speed. The key ingredient that constitutes an evaluation and not any other sort of description is its use as a stand-in for (or operationalization of) a \textit{value}. 

A long line of scholars, especially in education and policy contexts, have developed theories and professional standards concerned with evaluating programs \cite{stufflebeam1985analysis,guba1989fourth,scriven1999fine,alkin2004evaluation,dahler2011evaluation,mertens2018program}. A shared emphasis seems to be that evaluations should not unquestioningly adopt the goals of program facilitators but instead take a broader societal view. Such emphasis can be found, for example, in pushes for \textit{stakeholder-based} approaches to evaluation \cite{guba1989fourth}. A similar theme from literature on program evaluation and auditing is the need for \textit{third-party} or external oversight in high-stakes decision-making systems \cite{costanza2022audits,raji2022outsider, power1997audit}. These works make clear that internal auditing and self-evaluation often fall short in settings where firms behave in ways counter to society's interests.

We use the word evaluation to highlight that the appropriateness of strategic behaviors is tied to questions of value: Only by understanding the values underlying a particular measurement can we conclude that a (strategic) behavioral response is appropriate or inappropriate. 

\section{Conclusion}
\label{sec:conclusion}

As machine learning and mechanism design expand into high-stakes social domains, they increasingly play a role in the creation of decision rules that affect people's lives.  In cases where individuals respond strategically to these rules, it can be tempting to categorize these behaviors as gaming or cheating.
This paper puts forward an expanded view of evaluative systems, where the decision subject isn't the only strategic actor who deserves social or ethical scrutiny. In hiring settings, for example, it is often the strategic methods and norms used to evaluate candidates that explain behavioral responses from decision subjects. In settings with grade inflation, schools and students align their interests and strategically behave in a way that undermines a broader societal measure of interest. Taking a normative perspective, we find that the moral standing of strategic behaviors often depends on the ways those behaviors are evaluated and motivated. We argue that questions about whether decision subjects are seen as `gaming the system' need to be viewed in light of a parallel set of questions about the interests of the evaluating institution. Our expanded (three-player) model of evaluation is able to shed greater light on a variety of scenarios where certain strategies are either in line with or at odds with the interests of others.

There are a number of promising directions for future work. Though we use a three-player model to illustrate external social considerations in evaluation systems, there is no reason to stop at three. Many systems of evaluation have a recursive structure, where each evaluator might need its own third-party oversight mechanism, creating a larger vertical hierarchy of participants. In addition to this type of vertical expansion of our model, we would welcome work aimed at disentangling the bundle of values we describe as `societal interests'---that is, a horizontal expansion in the sets of values that are juxtaposed against the actions of the evaluator. Delineating the norms and standards behind evaluations is a complex, context-dependent, and political undertaking with the potential to affect life-prospects in significant ways; failing to wrestle with this complexity may result in unfairly and illegitimately placing a thumb on the scale in favor of one or more of the stakeholders.

\begin{acks}
The authors would like to thank the members of the AI, Policy and Practice group (AIPP) at Cornell University and the Digital Life Initiative (DLI) at Cornell Tech for their feedback. In particular, we thank Solon Barocas, Smitha Milli, Kenny Peng, and Malte Ziewitz for illuminating conversations, suggestions and remarks.

The work is supported in part by a grant from the John D. and Catherine T. MacArthur Foundation. Ben Laufer is additionally supported by a LinkedIn-Bowers CIS PhD Fellowship, a doctoral fellowship from DLI, and a SaTC NSF grant CNS-1704527. Jon Kleinberg is additionally supported by a Vannevar Bush Faculty Fellowship and a grant from the Simons Foundation. Helen Nissenbaum is also supported by a SaTC NSF grant CNS-1801501.
\end{acks}

\bibliographystyle{ACM-Reference-Format}
\bibliography{sample-base}


\begin{thebibliography}{55}


\ifx \showCODEN    \undefined \def \showCODEN     #1{\unskip}     \fi
\ifx \showDOI      \undefined \def \showDOI       #1{#1}\fi
\ifx \showISBNx    \undefined \def \showISBNx     #1{\unskip}     \fi
\ifx \showISBNxiii \undefined \def \showISBNxiii  #1{\unskip}     \fi
\ifx \showISSN     \undefined \def \showISSN      #1{\unskip}     \fi
\ifx \showLCCN     \undefined \def \showLCCN      #1{\unskip}     \fi
\ifx \shownote     \undefined \def \shownote      #1{#1}          \fi
\ifx \showarticletitle \undefined \def \showarticletitle #1{#1}   \fi
\ifx \showURL      \undefined \def \showURL       {\relax}        \fi
\providecommand\bibfield[2]{#2}
\providecommand\bibinfo[2]{#2}
\providecommand\natexlab[1]{#1}
\providecommand\showeprint[2][]{arXiv:#2}

\bibitem[Abebe and Goldner(2018)]%
        {abebe2018mechanism}
\bibfield{author}{\bibinfo{person}{Rediet Abebe} {and} \bibinfo{person}{Kira
  Goldner}.} \bibinfo{year}{2018}\natexlab{}.
\newblock \showarticletitle{Mechanism design for social good}.
\newblock \bibinfo{journal}{\emph{AI Matters}} \bibinfo{volume}{4},
  \bibinfo{number}{3} (\bibinfo{year}{2018}), \bibinfo{pages}{27--34}.
\newblock


\bibitem[Alkin and Christie(2004)]%
        {alkin2004evaluation}
\bibfield{author}{\bibinfo{person}{Marvin~C Alkin} {and}
  \bibinfo{person}{Christina~A Christie}.} \bibinfo{year}{2004}\natexlab{}.
\newblock \showarticletitle{An evaluation theory tree}.
\newblock \bibinfo{journal}{\emph{Evaluation roots: Tracing theorists’ views
  and influences}} \bibinfo{volume}{2}, \bibinfo{number}{19}
  (\bibinfo{year}{2004}), \bibinfo{pages}{12--65}.
\newblock


\bibitem[Anderson(2022)]%
        {anderson2022asu}
\bibfield{author}{\bibinfo{person}{Nicole~Almond Anderson}.}
  \bibinfo{year}{2022}\natexlab{}.
\newblock \bibinfo{title}{ASU Law Program helps women master the golf club as a
  business tool}.
\newblock
\newblock
\urldef\tempurl%
\url{https://attorneyatlawmagazine.com/law-school/asu/asu-law-program-helps-women-master-the-golf-club-as-a-business-tool}
\showURL{%
\tempurl}


\bibitem[Athletics(2020)]%
        {world-athletics-track-surface}
\bibfield{author}{\bibinfo{person}{World Athletics}.}
  \bibinfo{year}{2020}\natexlab{}.
\newblock \bibinfo{title}{Certification System Procedures}.
\newblock
\newblock
\newblock
\shownote{Online at www.worldathletics.org}.


\bibitem[Bambauer and Zarsky(2018)]%
        {bambauer2018algorithm}
\bibfield{author}{\bibinfo{person}{Jane Bambauer} {and} \bibinfo{person}{Tal
  Zarsky}.} \bibinfo{year}{2018}\natexlab{}.
\newblock \showarticletitle{The algorithm game}.
\newblock \bibinfo{journal}{\emph{Notre Dame L. Rev.}}  \bibinfo{volume}{94}
  (\bibinfo{year}{2018}), \bibinfo{pages}{1}.
\newblock


\bibitem[Barabas et~al\mbox{.}(2020)]%
        {barabas2020studying}
\bibfield{author}{\bibinfo{person}{Chelsea Barabas}, \bibinfo{person}{Colin
  Doyle}, \bibinfo{person}{JB Rubinovitz}, {and} \bibinfo{person}{Karthik
  Dinakar}.} \bibinfo{year}{2020}\natexlab{}.
\newblock \showarticletitle{Studying up: reorienting the study of algorithmic
  fairness around issues of power}. In \bibinfo{booktitle}{\emph{Proceedings of
  the 2020 Conference on Fairness, Accountability, and Transparency}}.
  \bibinfo{publisher}{ACM}, \bibinfo{address}{Barcelona, Spain},
  \bibinfo{pages}{167--176}.
\newblock


\bibitem[Bogen and Rieke(2018)]%
        {bogen2018help}
\bibfield{author}{\bibinfo{person}{Miranda Bogen} {and} \bibinfo{person}{Aaron
  Rieke}.} \bibinfo{year}{2018}\natexlab{}.
\newblock \showarticletitle{Help wanted: An examination of hiring algorithms,
  equity, and bias}.
\newblock \bibinfo{journal}{\emph{Upturn.org}}  \bibinfo{volume}{0}
  (\bibinfo{year}{2018}).
\newblock


\bibitem[Bok(1979)]%
        {bok1999lying}
\bibfield{author}{\bibinfo{person}{Sissela Bok}.}
  \bibinfo{year}{1979}\natexlab{}.
\newblock \bibinfo{booktitle}{\emph{Lying: Moral choice in public and private
  life}}.
\newblock \bibinfo{publisher}{Vintage}, \bibinfo{address}{New York}.
\newblock


\bibitem[Bourdieu(1987)]%
        {bourdieu1987distinction}
\bibfield{author}{\bibinfo{person}{Pierre Bourdieu}.}
  \bibinfo{year}{1987}\natexlab{}.
\newblock \bibinfo{booktitle}{\emph{Distinction: A social critique of the
  judgement of taste}}.
\newblock \bibinfo{publisher}{Harvard university press},
  \bibinfo{address}{Cambridge, MA, USA}.
\newblock


\bibitem[Caesar(2017)]%
        {wired-two-hour-marathon}
\bibfield{author}{\bibinfo{person}{Ed Caesar}.}
  \bibinfo{year}{2017}\natexlab{}.
\newblock \showarticletitle{The Epic Untold Story of Nike’s (Almost) Perfect
  Marathon}.
\newblock \bibinfo{journal}{\emph{Wired}}  \bibinfo{volume}{0}
  (\bibinfo{date}{June} \bibinfo{year}{2017}).
\newblock


\bibitem[Chen et~al\mbox{.}(2018)]%
        {chen2018strategyproof}
\bibfield{author}{\bibinfo{person}{Yiling Chen}, \bibinfo{person}{Chara
  Podimata}, \bibinfo{person}{Ariel~D Procaccia}, {and} \bibinfo{person}{Nisarg
  Shah}.} \bibinfo{year}{2018}\natexlab{}.
\newblock \showarticletitle{Strategyproof linear regression in high
  dimensions}. In \bibinfo{booktitle}{\emph{Proceedings of the 2018 ACM
  Conference on Economics and Computation}}. \bibinfo{publisher}{ACM},
  \bibinfo{address}{Ithaca, NY, USA}, \bibinfo{pages}{9--26}.
\newblock


\bibitem[Cofone and Strandburg(2019)]%
        {cofone2019strategic}
\bibfield{author}{\bibinfo{person}{Ignacio~N Cofone} {and}
  \bibinfo{person}{Katherine~J Strandburg}.} \bibinfo{year}{2019}\natexlab{}.
\newblock \showarticletitle{Strategic games and algorithmic secrecy}.
\newblock \bibinfo{journal}{\emph{McGill Law Journal}} \bibinfo{volume}{64},
  \bibinfo{number}{4} (\bibinfo{year}{2019}), \bibinfo{pages}{623--663}.
\newblock


\bibitem[Costanza-Chock et~al\mbox{.}(2022)]%
        {costanza2022audits}
\bibfield{author}{\bibinfo{person}{Sasha Costanza-Chock},
  \bibinfo{person}{Inioluwa~Deborah Raji}, {and} \bibinfo{person}{Joy
  Buolamwini}.} \bibinfo{year}{2022}\natexlab{}.
\newblock \showarticletitle{Who Audits the Auditors? Recommendations from a
  field scan of the algorithmic auditing ecosystem}. In
  \bibinfo{booktitle}{\emph{2022 ACM Conference on Fairness, Accountability,
  and Transparency}}. \bibinfo{publisher}{ACM}, \bibinfo{address}{Seoul,
  Korea}, \bibinfo{pages}{1571--1583}.
\newblock


\bibitem[Crumbley et~al\mbox{.}(2010)]%
        {crumbley2010ethical}
\bibfield{author}{\bibinfo{person}{Donald~Larry Crumbley},
  \bibinfo{person}{Ronald~E Flinn}, {and} \bibinfo{person}{Kenneth~J
  Reichelt}.} \bibinfo{year}{2010}\natexlab{}.
\newblock \showarticletitle{What is ethical about grade inflation and
  coursework deflation?}
\newblock \bibinfo{journal}{\emph{Journal of Academic Ethics}}
  \bibinfo{volume}{8} (\bibinfo{year}{2010}), \bibinfo{pages}{187--197}.
\newblock


\bibitem[Dahler-Larsen(2011)]%
        {dahler2011evaluation}
\bibfield{author}{\bibinfo{person}{Peter Dahler-Larsen}.}
  \bibinfo{year}{2011}\natexlab{}.
\newblock \bibinfo{booktitle}{\emph{The evaluation society}}.
\newblock \bibinfo{publisher}{Stanford University Press},
  \bibinfo{address}{Stanford, CA, USA}.
\newblock


\bibitem[Finefter-Rosenbluh and Levinson(2015)]%
        {finefter2015wrong}
\bibfield{author}{\bibinfo{person}{Ilana Finefter-Rosenbluh} {and}
  \bibinfo{person}{Meira Levinson}.} \bibinfo{year}{2015}\natexlab{}.
\newblock \showarticletitle{What is wrong with grade inflation (if anything)?}
\newblock \bibinfo{journal}{\emph{Philosophical Inquiry in Education}}
  \bibinfo{volume}{23}, \bibinfo{number}{1} (\bibinfo{year}{2015}),
  \bibinfo{pages}{3--21}.
\newblock


\bibitem[Finocchiaro et~al\mbox{.}(2021)]%
        {finocchiaro2021bridging}
\bibfield{author}{\bibinfo{person}{Jessie Finocchiaro}, \bibinfo{person}{Roland
  Maio}, \bibinfo{person}{Faidra Monachou}, \bibinfo{person}{Gourab~K Patro},
  \bibinfo{person}{Manish Raghavan}, \bibinfo{person}{Ana-Andreea Stoica},
  {and} \bibinfo{person}{Stratis Tsirtsis}.} \bibinfo{year}{2021}\natexlab{}.
\newblock \showarticletitle{Bridging machine learning and mechanism design
  towards algorithmic fairness}. In \bibinfo{booktitle}{\emph{Proceedings of
  the 2021 ACM Conference on Fairness, Accountability, and Transparency}}.
  \bibinfo{publisher}{ACM}, \bibinfo{address}{Virtual},
  \bibinfo{pages}{489--503}.
\newblock


\bibitem[Grossman and Hart(1992)]%
        {grossman1992analysis}
\bibfield{author}{\bibinfo{person}{Sanford~J Grossman} {and}
  \bibinfo{person}{Oliver~D Hart}.} \bibinfo{year}{1992}\natexlab{}.
\newblock \bibinfo{booktitle}{\emph{An analysis of the principal-agent
  problem}}.
\newblock \bibinfo{publisher}{Springer}, \bibinfo{address}{New York}.
\newblock


\bibitem[Guba and Lincoln(1989)]%
        {guba1989fourth}
\bibfield{author}{\bibinfo{person}{Egon~G Guba} {and} \bibinfo{person}{Yvonna~S
  Lincoln}.} \bibinfo{year}{1989}\natexlab{}.
\newblock \bibinfo{booktitle}{\emph{Fourth generation evaluation}}.
\newblock \bibinfo{publisher}{Sage}, \bibinfo{address}{Newbury Park}.
\newblock


\bibitem[Hardt et~al\mbox{.}(2016)]%
        {hardt2016strategic}
\bibfield{author}{\bibinfo{person}{Moritz Hardt}, \bibinfo{person}{Nimrod
  Megiddo}, \bibinfo{person}{Christos Papadimitriou}, {and}
  \bibinfo{person}{Mary Wootters}.} \bibinfo{year}{2016}\natexlab{}.
\newblock \showarticletitle{Strategic classification}. In
  \bibinfo{booktitle}{\emph{Proceedings of the 2016 ACM conference on
  innovations in theoretical computer science}}. \bibinfo{publisher}{ACM},
  \bibinfo{address}{Cambridge, MA, USA}, \bibinfo{pages}{111--122}.
\newblock


\bibitem[Harris et~al\mbox{.}(2021)]%
        {harris2021stateful}
\bibfield{author}{\bibinfo{person}{Keegan Harris}, \bibinfo{person}{Hoda
  Heidari}, {and} \bibinfo{person}{Steven~Z Wu}.}
  \bibinfo{year}{2021}\natexlab{}.
\newblock \showarticletitle{Stateful strategic regression}.
\newblock \bibinfo{journal}{\emph{Advances in Neural Information Processing
  Systems}}  \bibinfo{volume}{34} (\bibinfo{year}{2021}),
  \bibinfo{pages}{28728--28741}.
\newblock


\bibitem[Hitzig(2020)]%
        {hitzig2020normative}
\bibfield{author}{\bibinfo{person}{Zo{\"e} Hitzig}.}
  \bibinfo{year}{2020}\natexlab{}.
\newblock \showarticletitle{The normative gap: mechanism design and ideal
  theories of justice}.
\newblock \bibinfo{journal}{\emph{Economics \& Philosophy}}
  \bibinfo{volume}{36}, \bibinfo{number}{3} (\bibinfo{year}{2020}),
  \bibinfo{pages}{407--434}.
\newblock


\bibitem[Holmstr{\"o}m(1999)]%
        {holmstrom1999managerial}
\bibfield{author}{\bibinfo{person}{Bengt Holmstr{\"o}m}.}
  \bibinfo{year}{1999}\natexlab{}.
\newblock \showarticletitle{Managerial incentive problems: A dynamic
  perspective}.
\newblock \bibinfo{journal}{\emph{The review of Economic studies}}
  \bibinfo{volume}{66}, \bibinfo{number}{1} (\bibinfo{year}{1999}),
  \bibinfo{pages}{169--182}.
\newblock


\bibitem[Hu and Chen(2018)]%
        {hu2018short}
\bibfield{author}{\bibinfo{person}{Lily Hu} {and} \bibinfo{person}{Yiling
  Chen}.} \bibinfo{year}{2018}\natexlab{}.
\newblock \showarticletitle{A short-term intervention for long-term fairness in
  the labor market}. In \bibinfo{booktitle}{\emph{Proceedings of the 2018 World
  Wide Web Conference}}. \bibinfo{publisher}{ACM}, \bibinfo{address}{Lyon,
  France}, \bibinfo{pages}{1389--1398}.
\newblock


\bibitem[Hu et~al\mbox{.}(2019)]%
        {hu2019disparate}
\bibfield{author}{\bibinfo{person}{Lily Hu}, \bibinfo{person}{Nicole
  Immorlica}, {and} \bibinfo{person}{Jennifer~Wortman Vaughan}.}
  \bibinfo{year}{2019}\natexlab{}.
\newblock \showarticletitle{The disparate effects of strategic manipulation}.
  In \bibinfo{booktitle}{\emph{Proceedings of the Conference on Fairness,
  Accountability, and Transparency}}. \bibinfo{publisher}{ACM},
  \bibinfo{address}{Atlanta, GA, USA}, \bibinfo{pages}{259--268}.
\newblock


\bibitem[Joyner(1991)]%
        {joyner1991modeling}
\bibfield{author}{\bibinfo{person}{Michael~J Joyner}.}
  \bibinfo{year}{1991}\natexlab{}.
\newblock \showarticletitle{Modeling: optimal marathon performance on the basis
  of physiological factors}.
\newblock \bibinfo{journal}{\emph{Journal of applied physiology}}
  \bibinfo{volume}{70}, \bibinfo{number}{2} (\bibinfo{year}{1991}),
  \bibinfo{pages}{683--687}.
\newblock


\bibitem[Kleinberg and Raghavan(2020)]%
        {kleinberg2020classifiers}
\bibfield{author}{\bibinfo{person}{Jon Kleinberg} {and} \bibinfo{person}{Manish
  Raghavan}.} \bibinfo{year}{2020}\natexlab{}.
\newblock \showarticletitle{How do classifiers induce agents to invest effort
  strategically?}
\newblock \bibinfo{journal}{\emph{ACM Transactions on Economics and Computation
  (TEAC)}} \bibinfo{volume}{8}, \bibinfo{number}{4} (\bibinfo{year}{2020}),
  \bibinfo{pages}{1--23}.
\newblock


\bibitem[Kohli(2014)]%
        {kohli2014princeton}
\bibfield{author}{\bibinfo{person}{Sonali Kohli}.}
  \bibinfo{year}{2014}\natexlab{}.
\newblock \bibinfo{title}{Princeton is giving up ground in its fight against
  grade inflation}.
\newblock
\newblock
\urldef\tempurl%
\url{https://qz.com/277288/princeton-is-giving-up-ground-in-its-fight-against-grade-inflation}
\showURL{%
\tempurl}


\bibitem[Kohn(2019)]%
        {kohn2019why}
\bibfield{author}{\bibinfo{person}{Alfie Kohn}.}
  \bibinfo{year}{2019}\natexlab{}.
\newblock \bibinfo{title}{Why Can't Everyone Get A's?}
\newblock
\newblock
\urldef\tempurl%
\url{https://www.nytimes.com/2019/06/15/opinion/sunday/schools-testing-ranking.html}
\showURL{%
\tempurl}


\bibitem[Lanning and Perkins(1995)]%
        {lanning1995grade}
\bibfield{author}{\bibinfo{person}{Wayne Lanning} {and} \bibinfo{person}{Peggy
  Perkins}.} \bibinfo{year}{1995}\natexlab{}.
\newblock \showarticletitle{Grade inflation: A consideration of additional
  causes}.
\newblock \bibinfo{journal}{\emph{Journal of Instructional Psychology}}
  \bibinfo{volume}{22}, \bibinfo{number}{2} (\bibinfo{year}{1995}),
  \bibinfo{pages}{163}.
\newblock


\bibitem[Liu et~al\mbox{.}(2022)]%
        {liu2022strategic}
\bibfield{author}{\bibinfo{person}{Lydia~T Liu}, \bibinfo{person}{Nikhil Garg},
  {and} \bibinfo{person}{Christian Borgs}.} \bibinfo{year}{2022}\natexlab{}.
\newblock \showarticletitle{Strategic ranking}. In
  \bibinfo{booktitle}{\emph{International Conference on Artificial Intelligence
  and Statistics}}. \bibinfo{publisher}{PMLR}, \bibinfo{address}{Virtual},
  \bibinfo{pages}{2489--2518}.
\newblock


\bibitem[Mertens and Wilson(2018)]%
        {mertens2018program}
\bibfield{author}{\bibinfo{person}{Donna~M Mertens} {and}
  \bibinfo{person}{Amy~T Wilson}.} \bibinfo{year}{2018}\natexlab{}.
\newblock \bibinfo{booktitle}{\emph{Program evaluation theory and practice}}.
\newblock \bibinfo{publisher}{Guilford Publications}, \bibinfo{address}{New
  York}.
\newblock


\bibitem[Miller et~al\mbox{.}(2020)]%
        {miller2020strategic}
\bibfield{author}{\bibinfo{person}{John Miller}, \bibinfo{person}{Smitha
  Milli}, {and} \bibinfo{person}{Moritz Hardt}.}
  \bibinfo{year}{2020}\natexlab{}.
\newblock \showarticletitle{Strategic classification is causal modeling in
  disguise}. In \bibinfo{booktitle}{\emph{International Conference on Machine
  Learning}}. \bibinfo{publisher}{PMLR}, \bibinfo{address}{Virtual},
  \bibinfo{pages}{6917--6926}.
\newblock


\bibitem[Milli et~al\mbox{.}(2019)]%
        {milli2019social}
\bibfield{author}{\bibinfo{person}{Smitha Milli}, \bibinfo{person}{John
  Miller}, \bibinfo{person}{Anca~D Dragan}, {and} \bibinfo{person}{Moritz
  Hardt}.} \bibinfo{year}{2019}\natexlab{}.
\newblock \showarticletitle{The social cost of strategic classification}. In
  \bibinfo{booktitle}{\emph{Proceedings of the Conference on Fairness,
  Accountability, and Transparency}}. \bibinfo{publisher}{ACM},
  \bibinfo{address}{Atlanta, GA, USA}, \bibinfo{pages}{230--239}.
\newblock


\bibitem[Narayanan(2022)]%
        {narayanan2022limits}
\bibfield{author}{\bibinfo{person}{Arvind Narayanan}.}
  \bibinfo{year}{2022}\natexlab{}.
\newblock \bibinfo{title}{The limits of the quantitative approach to
  discrimination}.
\newblock
\newblock


\bibitem[Panja(2021)]%
        {nyt-olympic-track}
\bibfield{author}{\bibinfo{person}{Tariq Panja}.}
  \bibinfo{year}{2021}\natexlab{}.
\newblock \showarticletitle{A Track Built for Speed Is Already Producing
  Records}.
\newblock \bibinfo{journal}{\emph{New York Times}}  \bibinfo{volume}{0}
  (\bibinfo{date}{2 August} \bibinfo{year}{2021}).
\newblock


\bibitem[Petre et~al\mbox{.}(2019)]%
        {petre2019gaming}
\bibfield{author}{\bibinfo{person}{Caitlin Petre}, \bibinfo{person}{Brooke~Erin
  Duffy}, {and} \bibinfo{person}{Emily Hund}.} \bibinfo{year}{2019}\natexlab{}.
\newblock \showarticletitle{“Gaming the system”: Platform paternalism and
  the politics of algorithmic visibility}.
\newblock \bibinfo{journal}{\emph{Social Media+ Society}} \bibinfo{volume}{5},
  \bibinfo{number}{4} (\bibinfo{year}{2019}),
  \bibinfo{pages}{2056305119879995}.
\newblock


\bibitem[Power(1997)]%
        {power1997audit}
\bibfield{author}{\bibinfo{person}{Michael Power}.}
  \bibinfo{year}{1997}\natexlab{}.
\newblock \bibinfo{booktitle}{\emph{The audit society: Rituals of
  verification}}.
\newblock \bibinfo{publisher}{OUP Oxford}, \bibinfo{address}{Oxford}.
\newblock


\bibitem[Quillian et~al\mbox{.}(2020)]%
        {quillian2020evidence}
\bibfield{author}{\bibinfo{person}{Lincoln Quillian}, \bibinfo{person}{John~J
  Lee}, {and} \bibinfo{person}{Mariana Oliver}.}
  \bibinfo{year}{2020}\natexlab{}.
\newblock \showarticletitle{Evidence from field experiments in hiring shows
  substantial additional racial discrimination after the callback}.
\newblock \bibinfo{journal}{\emph{Social Forces}} \bibinfo{volume}{99},
  \bibinfo{number}{2} (\bibinfo{year}{2020}), \bibinfo{pages}{732--759}.
\newblock


\bibitem[Raghavan et~al\mbox{.}(2020)]%
        {raghavan2020mitigating}
\bibfield{author}{\bibinfo{person}{Manish Raghavan}, \bibinfo{person}{Solon
  Barocas}, \bibinfo{person}{Jon Kleinberg}, {and} \bibinfo{person}{Karen
  Levy}.} \bibinfo{year}{2020}\natexlab{}.
\newblock \showarticletitle{Mitigating bias in algorithmic hiring: Evaluating
  claims and practices}. In \bibinfo{booktitle}{\emph{Proceedings of the 2020
  conference on fairness, accountability, and transparency}}.
  \bibinfo{publisher}{ACM}, \bibinfo{address}{Barcelona, Spain},
  \bibinfo{pages}{469--481}.
\newblock


\bibitem[Raji et~al\mbox{.}(2022)]%
        {raji2022outsider}
\bibfield{author}{\bibinfo{person}{Inioluwa~Deborah Raji},
  \bibinfo{person}{Peggy Xu}, \bibinfo{person}{Colleen Honigsberg}, {and}
  \bibinfo{person}{Daniel Ho}.} \bibinfo{year}{2022}\natexlab{}.
\newblock \showarticletitle{Outsider oversight: Designing a third party audit
  ecosystem for ai governance}. In \bibinfo{booktitle}{\emph{Proceedings of the
  2022 AAAI/ACM Conference on AI, Ethics, and Society}}.
  \bibinfo{publisher}{AAAI/ACM}, \bibinfo{address}{Oxford, United Kingdom},
  \bibinfo{pages}{557--571}.
\newblock


\bibitem[Ray(2019)]%
        {ray2019theory}
\bibfield{author}{\bibinfo{person}{Victor Ray}.}
  \bibinfo{year}{2019}\natexlab{}.
\newblock \showarticletitle{A theory of racialized organizations}.
\newblock \bibinfo{journal}{\emph{American Sociological Review}}
  \bibinfo{volume}{84}, \bibinfo{number}{1} (\bibinfo{year}{2019}),
  \bibinfo{pages}{26--53}.
\newblock


\bibitem[Rivera(2016)]%
        {rivera2016pedigree}
\bibfield{author}{\bibinfo{person}{Lauren~A Rivera}.}
  \bibinfo{year}{2016}\natexlab{}.
\newblock \showarticletitle{Pedigree}.
\newblock In \bibinfo{booktitle}{\emph{Pedigree}}.
  \bibinfo{publisher}{Princeton University Press}, \bibinfo{address}{Princeton,
  NJ, USA}.
\newblock


\bibitem[Rojstaczer and Healy(2012)]%
        {rojstaczer2012ordinary}
\bibfield{author}{\bibinfo{person}{Stuart Rojstaczer} {and}
  \bibinfo{person}{Christopher Healy}.} \bibinfo{year}{2012}\natexlab{}.
\newblock \showarticletitle{Where A is ordinary: The evolution of American
  college and university grading, 1940-2009}.
\newblock \bibinfo{journal}{\emph{Teachers College Record}}
  \bibinfo{volume}{114}, \bibinfo{number}{7} (\bibinfo{year}{2012}),
  \bibinfo{pages}{1--23}.
\newblock


\bibitem[Scriven(1991)]%
        {scriven1991evaluation}
\bibfield{author}{\bibinfo{person}{Michael Scriven}.}
  \bibinfo{year}{1991}\natexlab{}.
\newblock \bibinfo{booktitle}{\emph{Evaluation thesaurus}}.
\newblock \bibinfo{publisher}{Sage}, \bibinfo{address}{Newbury Park}.
\newblock


\bibitem[Scriven(1999)]%
        {scriven1999fine}
\bibfield{author}{\bibinfo{person}{Michael Scriven}.}
  \bibinfo{year}{1999}\natexlab{}.
\newblock \showarticletitle{The fine line between evaluation and explanation}.
\newblock \bibinfo{journal}{\emph{Research on social work practice}}
  \bibinfo{volume}{9}, \bibinfo{number}{4} (\bibinfo{year}{1999}),
  \bibinfo{pages}{521--524}.
\newblock


\bibitem[Scriven(2007)]%
        {scriven2007logic}
\bibfield{author}{\bibinfo{person}{Michael Scriven}.}
  \bibinfo{year}{2007}\natexlab{}.
\newblock \showarticletitle{The logic of evaluation}.
\newblock \bibinfo{journal}{\emph{Dissensus and the Search for Common Ground}}
  \bibinfo{volume}{1} (\bibinfo{year}{2007}), \bibinfo{pages}{1--16}.
\newblock


\bibitem[Shavit et~al\mbox{.}(2020)]%
        {shavit2020causal}
\bibfield{author}{\bibinfo{person}{Yonadav Shavit}, \bibinfo{person}{Benjamin
  Edelman}, {and} \bibinfo{person}{Brian Axelrod}.}
  \bibinfo{year}{2020}\natexlab{}.
\newblock \showarticletitle{Causal strategic linear regression}. In
  \bibinfo{booktitle}{\emph{International Conference on Machine Learning}}.
  \bibinfo{publisher}{PMLR}, \bibinfo{address}{Virtual},
  \bibinfo{pages}{8676--8686}.
\newblock


\bibitem[Stufflebeam et~al\mbox{.}(1985)]%
        {stufflebeam1985analysis}
\bibfield{author}{\bibinfo{person}{Daniel~L Stufflebeam},
  \bibinfo{person}{Anthony~J Shinkfield}, \bibinfo{person}{Daniel~L
  Stufflebeam}, {and} \bibinfo{person}{Anthony~J Shinkfield}.}
  \bibinfo{year}{1985}\natexlab{}.
\newblock \showarticletitle{An analysis of alternative approaches to
  evaluation}.
\newblock \bibinfo{journal}{\emph{Systematic Evaluation: A Self-Instructional
  Guide to Theory and Practice}}  \bibinfo{volume}{1} (\bibinfo{year}{1985}),
  \bibinfo{pages}{45--68}.
\newblock


\bibitem[Susser et~al\mbox{.}(2019a)]%
        {susser2019online}
\bibfield{author}{\bibinfo{person}{Daniel Susser}, \bibinfo{person}{Beate
  Roessler}, {and} \bibinfo{person}{Helen Nissenbaum}.}
  \bibinfo{year}{2019}\natexlab{a}.
\newblock \showarticletitle{Online manipulation: Hidden influences in a digital
  world}.
\newblock \bibinfo{journal}{\emph{Geo. L. Tech. Rev.}}  \bibinfo{volume}{4}
  (\bibinfo{year}{2019}), \bibinfo{pages}{1--45}.
\newblock


\bibitem[Susser et~al\mbox{.}(2019b)]%
        {susser2019technology}
\bibfield{author}{\bibinfo{person}{Daniel Susser}, \bibinfo{person}{Beate
  Roessler}, {and} \bibinfo{person}{Helen Nissenbaum}.}
  \bibinfo{year}{2019}\natexlab{b}.
\newblock \showarticletitle{Technology, autonomy, and manipulation}.
\newblock \bibinfo{journal}{\emph{Internet Policy Review}} \bibinfo{volume}{8},
  \bibinfo{number}{2} (\bibinfo{year}{2019}), \bibinfo{pages}{1--22}.
\newblock


\bibitem[Vecchione et~al\mbox{.}(2021)]%
        {vecchione2021algorithmic}
\bibfield{author}{\bibinfo{person}{Briana Vecchione}, \bibinfo{person}{Karen
  Levy}, {and} \bibinfo{person}{Solon Barocas}.}
  \bibinfo{year}{2021}\natexlab{}.
\newblock \showarticletitle{Algorithmic auditing and social justice: Lessons
  from the history of audit studies}.
\newblock In \bibinfo{booktitle}{\emph{Equity and Access in Algorithms,
  Mechanisms, and Optimization}}. \bibinfo{publisher}{ACM},
  \bibinfo{address}{Virtual}, \bibinfo{pages}{1--9}.
\newblock


\bibitem[Viljoen et~al\mbox{.}(2021)]%
        {viljoen2021design}
\bibfield{author}{\bibinfo{person}{Salom{\'e} Viljoen}, \bibinfo{person}{Jake
  Goldenfein}, {and} \bibinfo{person}{Lee McGuigan}.}
  \bibinfo{year}{2021}\natexlab{}.
\newblock \showarticletitle{Design choices: Mechanism design and platform
  capitalism}.
\newblock \bibinfo{journal}{\emph{Big data \& society}} \bibinfo{volume}{8},
  \bibinfo{number}{2} (\bibinfo{year}{2021}),
  \bibinfo{pages}{20539517211034312}.
\newblock


\bibitem[Ziewitz(2019)]%
        {ziewitz2019rethinking}
\bibfield{author}{\bibinfo{person}{Malte Ziewitz}.}
  \bibinfo{year}{2019}\natexlab{}.
\newblock \showarticletitle{Rethinking gaming: The ethical work of optimization
  in web search engines}.
\newblock \bibinfo{journal}{\emph{Social studies of science}}
  \bibinfo{volume}{49}, \bibinfo{number}{5} (\bibinfo{year}{2019}),
  \bibinfo{pages}{707--731}.
\newblock


\bibitem[Zrnic et~al\mbox{.}(2021)]%
        {zrnic2021leads}
\bibfield{author}{\bibinfo{person}{Tijana Zrnic}, \bibinfo{person}{Eric
  Mazumdar}, \bibinfo{person}{Shankar Sastry}, {and} \bibinfo{person}{Michael
  Jordan}.} \bibinfo{year}{2021}\natexlab{}.
\newblock \showarticletitle{Who Leads and Who Follows in Strategic
  Classification?}
\newblock \bibinfo{journal}{\emph{Advances in Neural Information Processing
  Systems}}  \bibinfo{volume}{34} (\bibinfo{year}{2021}),
  \bibinfo{pages}{15257--15269}.
\newblock


\end{thebibliography}

\appendix

\end{document}